\documentclass{article}
\usepackage{hiph-preprint}
\volnumber{22} \issuenumber{1} \edyear{2005}                             %|
\frompage{000} \topage{000}                                              %|
\recrevdate{27 October 2005}                                             %|
\def\rightmark{$e^+ e^-$ Pairs}
\def\leftmark{A. J. Baltz}
 
\title{Heavy Ion $e^+ e^-$ Pairs to All Orders in Z $\alpha$}
\authors{ 
{Anthony J. Baltz$^1$%
\index{One, A.} % Baltz,
\index{Two, A.} % to be inserted for use in the volume index
}\\[2.812mm]
{\normalsize
\hspace*{-8pt}$^1$ Brookhaven National Laboratory,\\ 
Upton, NY 11973, USA\\[0.2ex] 
}}
 
\abstract{The heavy ion cross section for continuum $e^+ e^-$ pair production
has been calculated to all orders in $Z \alpha$.  Comparison is made with
available
CERN SPS and RHIC STAR data.  Computed cross sections are found
to be reduced from perturbation theory with increasing charge of the colliding
heavy ions and for all energy and momentum regions investigated.  Au or Pb 
total cross sections are reduced by 28\% (SPS), 17\% (RHIC), and 11\% (LHC).
For very high energy ($E_{e^+}, E_{e^-} > 3$ GeV) forward pairs
at LHC the reduction from perturbation theory is a bit larger (17\%).  
Use of zero degree calorimeter triggering (and thus small impact parameter
weighting) makes impact parameter
representation of exact pair production useful.  Preliminary exact calculations
in the zero impact parameter limit show a much larger 
reduction from perturbation theory (about 40\%) at both RHIC and LHC.}

 
\makeindex
\begin{document}
 
\maketitle

\section{Introduction}
A two center light cone calculation of continuum pairs can be carried out
by exactly solving the
semi-classical Dirac equation\cite{sw,bm,sw2,er} for colliding $\delta$
function potentials
$$
V(\mbox{\boldmath $ \rho$},z,t)
=\delta(z - t) (1-\alpha_z) \Lambda^-(\mbox{\boldmath $ \rho$})
+\delta(z + t) (1+\alpha_z) \Lambda^+(\mbox{\boldmath $ \rho$}) 
$$
in the collider center of mass (lab) frame, where
$$
\Lambda^{\pm}(\mbox{\boldmath $ \rho$}) = - Z \alpha 
\ln {(\mbox{\boldmath $ \rho$} \pm {\bf b}/2)^2 \over (b/2)^2}.
$$

Baltz and McLerran\cite{bm} originally noted an agreement with
perturbation theory in the exact result.
Segev and Wells\cite{sw2} noted this agreement and also noted the scaling with
$Z_1^2 Z_2^2$ seen in SPS data for 160 GeV/c Pb ions on C, Al, Pa, Au
and for 200 Gev/c S ions on C, Al, Pa, Au.
The experimental group, Vane et al., summarized their data:
``Cross sections scale as the product of the squares of the projectile
and target nuclear charges\cite{vd}.''

On the other hand, photoproduction on a heavy target shows a negative
correction proportional to $Z^2$\cite{bethe}.
Several authors have argued that a correct regularization of the exact
Dirac equation amplitude should lead to Coulomb corrections\cite{serbo,lm}.  

\section{Cross Section with Higher Order Coulomb Corrections}
I have previously showed how a physical cutoff of the transverse potential
$\Lambda^{\pm}(\mbox{\boldmath $ \rho$})$
leads to Coulomb corrections\cite{ajb3} consistent with the Lee and Milstein
approximate analytical result\cite{lm}.  In this section I present recent
``exact'' Dirac equation calculations of total cross sections that make use
of this physical cutoff\cite{ajb4}.

Table \ref{tab1} shows the results of exact numerical calculations indicating
reductions from perturbation theory.  In addition, calculations of positron
transverse and longitudinal spectra show that the reductions persist to the
highest and lowest momentum values.

Note that the total cross section at CERN SPS energy is reduced from
perturbation theory by 28\%, a disagreement with the experimentally presented
perturbative scaling.  Nevertheless, given the difficulty of the SPS experiment
as described by the authors, I would argue that the apparent lack of Coulomb
corrections in the data
needs to be verified in other ultrarelativistic heavy ion experiments.

\begin{table}[hb]
\vspace*{-12pt}
\caption[]{Computer calculations compared with analytical formula results. 
$\gamma$ is defined for one of the ions in the frame of equal magnitude and
opposite direction velocities.  Total cross sections are expressed in barns.}
\label{tab1}\label{tabi}
\begin{tabular}{ccccc}
\hline\\[-10pt]
& & Exact & Perturb. & Difference \\
\hline\\[-10pt]
Pb + Au&Computer Evaluation\cite{ajb4}& 2670 & 3720 & -1050  \\
$\gamma = 9.2$&Racah Formula\cite{rac}&  &  3480&  \\
SPS&Lee-Milstein\cite{lm}& 3050 & 5120 & -2070  \\
&&&& \\
Au + Au&Computer Evaluation& 28,600 & 34,600 & -6,000  \\
$\gamma = 100$&Racah Formula&  &  34,200& \\
RHIC&Lee-Milstein& 34,100 & 42,500 & -8,400  \\
&Hencken, Trautman, Baur\cite{henck}&  & 34,000 &  \\
&&&& \\
Pb + Pb&Computer Evaluation& 199,000 & 224,000 & -25,000  \\
$\gamma = 2960$&Racah Formula&  &  226,000& \\
LHC&Lee-Milstein& 226,000 & 258,000 & -32,000  \\
\hline 
\end{tabular}
\end{table}
 
\section{RHIC STAR Data}
$e^+ e^-$ pair production accompanied by nuclear dissociation has been
measured by STAR.  Comparison with
perturbative QED calculations allowed
a limit to be set ``on higher-order corrections to the cross section,
$-0.5 \sigma_{QED} < \Delta \sigma < 0.2 \sigma_{QED}$,
at a 90\% confidence level\cite{star}.''

Calculations
in the STAR acceptance without dissociation provide an indication
of the relative difference between perturbation theory and the
exact result.  In the STAR acceptance the exact result is calculated to
be 17\% lower than perturbation theory.  This rough estimate,
$\Delta \sigma = -0.17 \sigma_{QED}$, is not excluded by STAR.

\section{Forward Pairs at LHC}
A sample numerical calculation has been performed using the
same method for $e^+ e^-$ production by Pb + Pb ions with cuts from a possible
detector setup at LHC.  With electron and positron energy E and angle $\theta$
in the range, $3 {\rm \ Gev} < {\rm E} < 20 {\rm \ GeV} $ and 
$ .00223 \ {\rm radians} <  \theta <  .00817 \ {\rm radians},$
the no
form factor perturbation theory cross section of 2.88 b is reduced by 18\%
to 2.36 b in an exact numerical calculation. 

If forward $e^+ e^-$ pairs are to
be employed for luminosity measurements at LHC, then it seems necessary to
consider the Coulomb corrections to the predicted cross sections.

\section{Probabilities at Small Impact Parameter: the Zero Impact Parameter
Limit}
Zero degree calorimeter (ZDC) triggering (e.g. for the STAR pair production)
weights smaller
parameter contributions: the probability at each impact parameter goes
as the product of the dissociation probabilty (ZDC) and the pair production
probability.  Pair production probability as a function of impact parameter is
needed to describe ZDC triggered events as was done for $\rho$ production at
STAR\cite{rho}.  I calculate the number weighted probability $P_T$
(or number operator), $P_T = \sum_{n=1}^{\infty} n P_n(b)$,
for producing $e^+ e^-$ pairs at some impact parameter $b$.
As a first step I compute the ${\bf b} = 0$ limit.

For Au + Au at RHIC the pertubation theory result is $P^0(0) = 1.64$,
the exact result $P(0) = .94$, or $P(0) = .57 P^0(0)$.  This limit may
have some relevance to the very high energy pairs measured by STAR which
necessarily come from relatively low impact parameters.  Again 
$-0.5 \sigma_{QED} < \Delta \sigma < 0.2 \sigma_{QED}$ from STAR
is not contrary the indications from the exact result in this limit.

For Pb + Pb at LHC the pertubation theory result is $P^0(0) = 4.07$,
and the exact result is $P(0) = 2.39 = .59 P^0(0)$.

\section{Conclusions}\label{concl}
A full numerical evaluation of the ``exact''
total cross section for $e^+ e^-$ production with gold or lead ions
shows reductions from perturbation theory of 28\% (SPS), 17\% (RHIC),
and 11\%(LHC).  Reductions are 43\% (RHIC), and 41\%(LHC) for $b = 0$.
For large Z no final momentum region was found in which there was
no reduction
or an insignificant reduction of the exact cross section.

\section{For the Future}\label{fut}
The predicted reduction of continuum $e^+e^-$ pair production from
$Z_A^2 Z_B^2$ scaling at higher $Z$ has never been observed experimentally.
An obvious suggestion for experiment is to compare electromagnetic $e^+e^-$
pair production in Au + Au with e.g. Ca + Ca at RHIC or LHC.

There is also a clear need to construct a computer program to
calculate P(b) exactly for $e^+e^-$ to all order in $Z \alpha$.
 
\section*{Acknowledgment}
 
This manuscript has been authored
under Contract No. DE-AC02-98CH10886 with the U. S. Department of Energy.

\vfill\eject
\end{document}